\documentclass[11pt,dvips]{article}

\usepackage{epsfig,times} 
\usepackage{picinpar}
\usepackage{wrapfig}
\usepackage{floatflt}
\makeatletter
\renewcommand{\section}{\@startsection{section}{1}{0in}
	{0.4\baselineskip}{0.1\baselineskip}{\Large\bf}}
\renewcommand{\subsection}{\@startsection{subsection}{2}{0in}
	{0.25\baselineskip}{-\baselineskip}{\large\bf}}
\renewcommand{\subsubsection}{\@startsection{subsubsection}{3}{0in}
	{0.1\baselineskip}{-\baselineskip}{\normalsize\bf}}
\makeatother
\begin{document}

%
\makeatletter\newcommand{\ps@icrc}{
\renewcommand{\@oddhead}{\slshape{HE.3.2.13}\hfil}}
\makeatother\thispagestyle{icrc}
%
%

\begin{center}
%
{\LARGE \bf Transition Radiation  Detector in MACRO }
\end{center}

\begin{center}
%
%
{\bf M.N. Mazziotta$^{1}$, for the MACRO Collaboration}\\
{\it $^{1}$Dipartimento di Fisica dell'Universit\'a and INFN Sezione
di Bari, via Amendola, 173, I-70126 Bari (Italy) }
\end{center}

\begin{center}
{\large \bf Abstract\\}
\end{center}
\vspace{-0.5ex}
%
%
The MACRO detector is located in the Gran Sasso Laboratory. MACRO's
overburden varies from $3150$ to $7000~hg/cm^2$. A transition radiation
detector (TRD) has been added to the MACRO detector in order to measure
the residual energy of muons entering MACRO, i.e. the energy they have
after passing through the Gran Sasso's rock overburden. The TRD 
consists of three identical modules with a total
horizontal area of $36~m^2$. The results presented here are referred to
single and double events in MACRO with one muon crossing one of the TRD
modules. Our data show that double muons are more energetic than single
ones, as predicted by the interaction models of primary cosmic rays with
the atmosphere.
\vspace{1ex}
%
%
\section{Introduction:}
\label{intro}
High energy underground muons are the remnants of the air showers
produced in the atmosphere by collisions of high energy cosmic ray
nuclei with air nuclei. Since muons are nearly stable and
have a small interaction cross section, they are called the
``penetrating charged component'' of cosmic rays. Thus, muons give the
dominant signal deep in the atmosphere and underground.
Muons carry information about the primary particle mass, the
primary energy spectrum and the inelastic cross section.
Underground muons also give information on energy loss in the rock.  

The analysis of the energy of muons detected deep underground is one of the
tools used for the indirect study of the interaction models of primary
cosmic rays.  As in all indirect measurements in cosmic ray physics, the
final interpretation is unavoidably dependent of the model adopted to
describe the secondary production and transport, and on the energy spectra
and chemical composition of primaries.
The energy loss of muons in the rock smears the information about 
primaries carried by the muons.
It is therefore crucial to find
physical observables which can be used to investigate the interaction
models besides the energy spectra and chemical composition of
the primary cosmic rays. However, it is very hard to disentangle the
interaction model from spectra and composition; thus in any discussion one
needs to take into account all those components, while dedicated
analyses (as depth intensity, decoherence) can be used to put some
constraints on the proprieties of the primary cosmic rays. 

In the present paper we describe a measurement of the
underground muon energy spectrum, carried out using a transition radiation
detector (TRD) in association with the MACRO apparatus. In this analysis
we use single and double muon events in MACRO, with one muon
crossing one of the TRD modules, in order to
investigate the all-particle energy spectrum of the primary cosmic
rays taking into account the energy loss of the muons in the rock
above the detector.

In a previous analysis (Ambrosio, et al, 1999),
which can be used as reference for the detector description and for the
analysis method, we used the single muons crossing the first TRD module.
In the present
analysis we use all the TRD modules, providing a large sample of
single muons and a sufficient number of double muons.

\section{Detector description:}
The MACRO detector is located in Hall B of the Gran Sasso 
Underground Laboratory. The laboratory is located at an average depth of 3700
$hg/cm^2$, with a minimum depth of 3150 $hg/cm^2$.
At these depths the residual energy differential distribution 
of the downgoing muons is estimated to be nearly 
flat up to 100 GeV and it then falls rapidly in the TeV region; the 
mean value is  a few hundred GeV. 
The TRD has been designed to explore the muon energy 
range of 100 GeV--1 TeV. Below 100 GeV there is no transition
radiation (TR) emission;
from 100 GeV to 1 TeV
the detector has a smoothly increasing response versus the muon energy.
For energies greater than 1 TeV, where the 
muon flux is estimated to be approximately $5\%$ of the total, 
the TR signal is saturated.

\begin{figwindow}[1,l,%
{\mbox{\epsfig{file=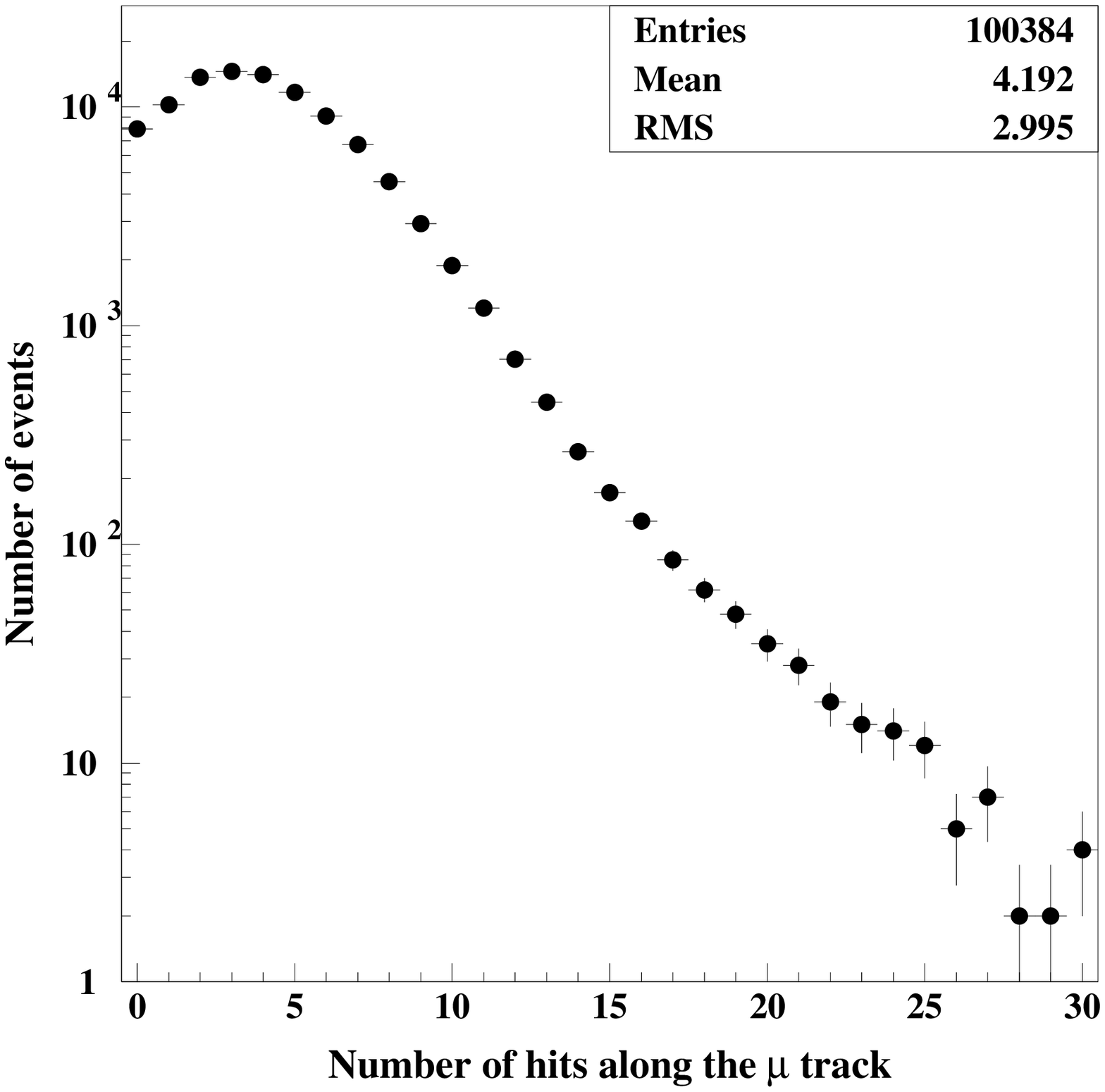,width=8.0cm,height=6.5cm}}},%
{Hit distribution for single muon tracks crossing the first TRD module.
}
\label{hit1}
\vspace{1ex}]
The MACRO TRD consists of three modules of 36 $m^2$ total horizontal
area. Each
module has an active volume of $6\times1.92\times1.7$ $m^3$
and contains 10 planes of 32 proportional tubes, $6$ meters long and with
a square cross section of $6\times6$ $cm^2$. 
These counters are laid close together 
between 11 Ethafoam radiator layers of 10 $cm$ height to form a large 
multiple layer TRD with reduced inefficient zones. The ionization loss 
and the X-rays of TR produced by muons are detected in the
proportional tubes filled with $Ar-CO_2$ mixture. A detailed 
description of the MACRO TRD is given in Barbarito
et al., 1995.

$~~~~~$ The first TRD module began to collect data in 1994, while
the second and the third TRD modules, which were put in acquisition
in 1996, are similar to the first module, but are equipped with a different
front-end electronics. This is the reason why the data samples of the
first and of the second and the third TRD modules need to be analyzed
separately.
Since the second module was built with the same structure as the
third module, a joint analysis of their data is possible.
\end{figwindow}

\section{Data Selection:}
For this analysis we have considered the data collected by the first
TRD module from April 1995 to January 1999, the data collected by the
second one from June 1997 to January 1999 and the data collected by the
third module from January 1997 to January 1999. We have analyzed two categories
of events:
``single muons", i.e. single events in MACRO crossing one TRD module; and
``double muons", i.e. double events in MACRO with only one muon
crossing one TRD module.
\begin{figwindow}[1,l,%
{\mbox{\epsfig{file=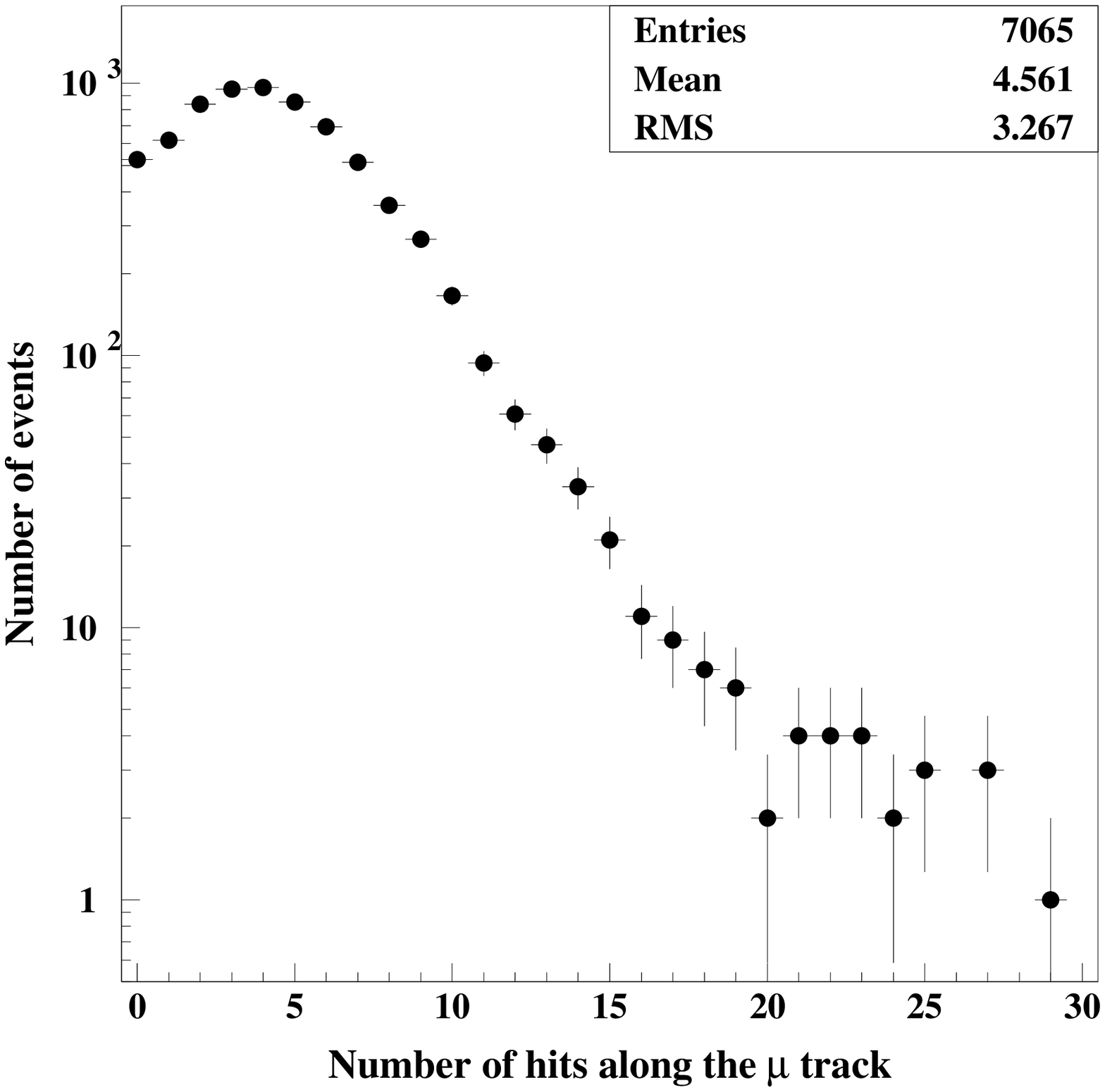,width=8.0cm,height=6.5cm}}},%
{TRD hit distribution in the first module for double muon tracks.
}
\label{hit2}
\vspace{1ex}]
Since the TRD calibration was performed 
for particles crossing the ten layers and at zenith angles below
$45^o$ (Barbarito et al., 1995), in this analysis only muons fulfilling these constraints have
been included.
The total number of 
hits in the track is evaluated by counting the number of TRD hits
along the straight 
line fitted to the track reconstructed by the MACRO detector. 
The total number of final events is about $2\cdot10^5$ for single
muons and about $13\cdot10^3$ for double muons.

$~~~~~$ In Figures \ref{hit1} and \ref{hit2} the distributions of the number 
of hits in the single muon and double muon tracks of the final event sample
in the first TRD module are shown.
It should be noted that the average value of the double muon distribution
is greater than the one of the corresponding single muon distribution. 
This means that the average energy of double muons is higher than the 
average energy of single muons. The ratio of the difference between
the average hit of double muons ($<hit>_{2\mu}$) and the average hit of
single muons ($<hit>_{1\mu}$) over the average hit of single muons are
$R_{hit}=(<hit>_{2\mu}-<hit>_{1\mu})/<hit>_{1\mu}=(8 \pm 1)\%$. 
This ratio is correlated to the energy ratio:
$R_{E}=(<E>_{2\mu}-<E>_{1\mu})/<E>_{1\mu}$.
\end{figwindow}

\section{Muon energy spectrum:}
In order to evaluate the local muon energy spectrum, we must 
take into account the TRD response function, which induces
some distortion of the ``true'' muon spectrum distribution.
The ``true'' distribution can be extracted from the measured one by an 
unfolding procedure that yields good results only if the 
response of the detector is correctly understood.
We have adopted an unfolding technique, developed according 
to Bayes' theorem, following the procedure described in D'Agostini, 1995
and Mazziotta, 1995.

\subsection{Detector simulation}
The distributions of the hits collected along a muon track at fixed
rock depth $h$ by the TRD and at 
a given zenithal and azimuthal angle, $N(k,\vartheta,\varphi)$, 
can be related to the residual energy 
distribution of muons, $N(E,\vartheta,\varphi)$, by: 
\begin{equation}
\label{eq:e1}
N(k,\vartheta,\varphi) = \sum_{j=1}^{n_E} p(k \mid E_j,\vartheta,\varphi) 
N(E_j,\vartheta,\varphi)
\end{equation}
where the detector response function, $p(k \mid E_j,\vartheta,\varphi)$,
represents the probability to observe k hits for a track of a given 
energy $E_j$ and at a given angle $\vartheta$ and $\varphi$.
The response function must contain both the detector acceptance and the 
event reconstruction efficiency. We have derived the response function
by simulating the MACRO behaviour using GEANT (Brun et al., 1992),
including the trigger efficiency simulation.
The simulation of the TRD was based on the test beam 
calibration data, taking into account the inefficiency 
of the proportional tubes.
A check of the response function of the TRD is obtained by using low energy
muons, namely stopping muons and muons with large scattering angles in
MACRO, which have energies of about $1-2~GeV$
(Ambrosio et al, 1999).

\subsection{Experimental data distributions}
The unfolding procedure has been applied separately to the TRD 
data samples of the first module, and of the second and the third modules,
starting with a trial spectrum assigned to the unfolded distribution
(D' Agostini, 1995; Mazziotta, 1995),
according to a local energy spectrum of muons at 4000 $hg/cm^2$
with a spectral index fixed at 3.7 given by (Lipari and Stanev, 1991):
\begin{equation}
\label{eq:e2}
N_o(E, h) \sim (E+\epsilon(1-e^{-\beta h}))^{-\alpha}
\end{equation}
The parameters are: $h=4000~hg/cm^{2}$, $\alpha=3.7$,
$\beta=0.383~10^{-3}~cm^2/hg$, $\epsilon=618~GeV$ and $E(GeV)$.

\begin{figwindow}[1,r,%
{\mbox{
\begin{tabular}{cc}
\epsfig{file=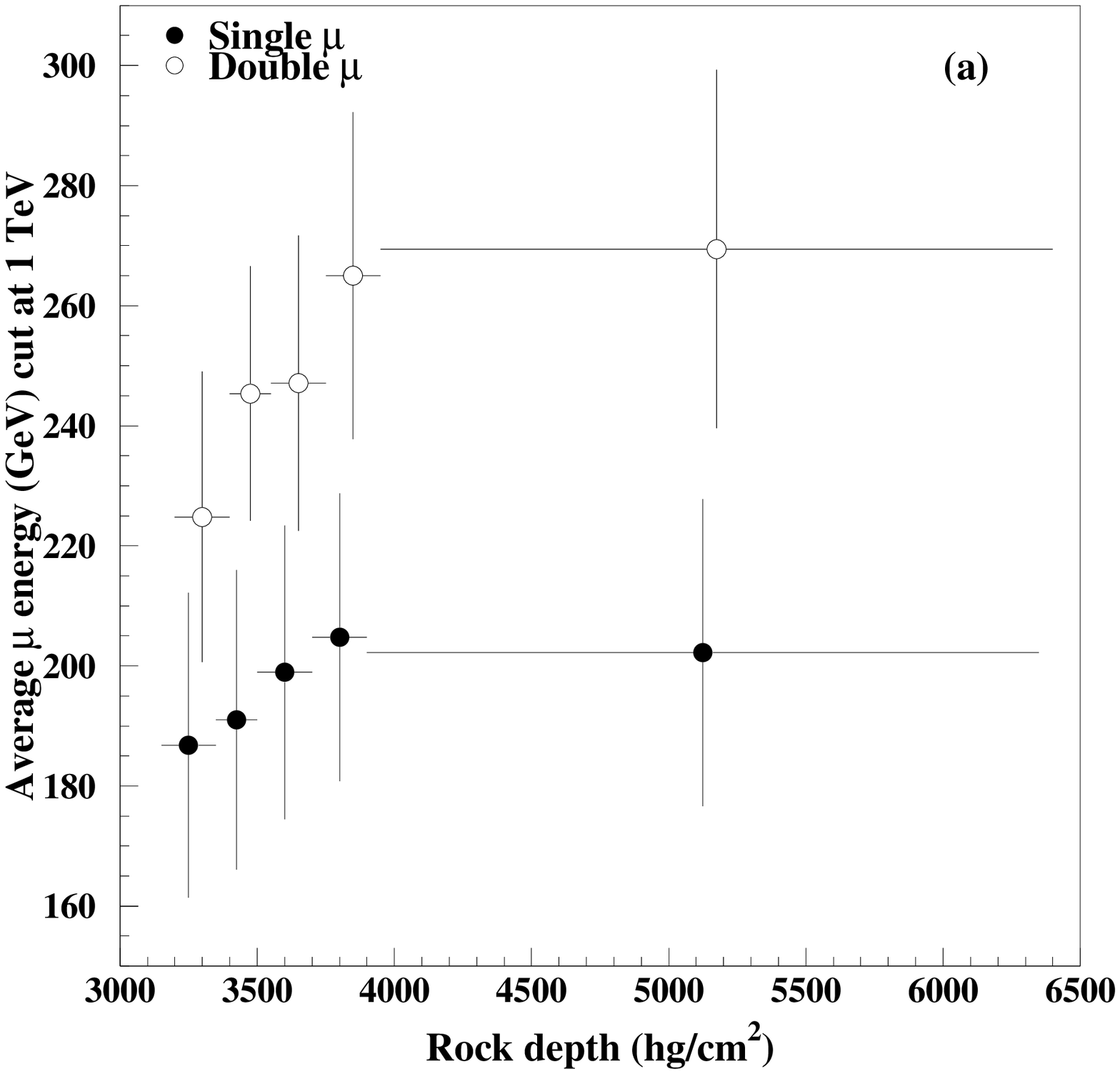,width=8.cm,height=6.5cm} &
\epsfig{file=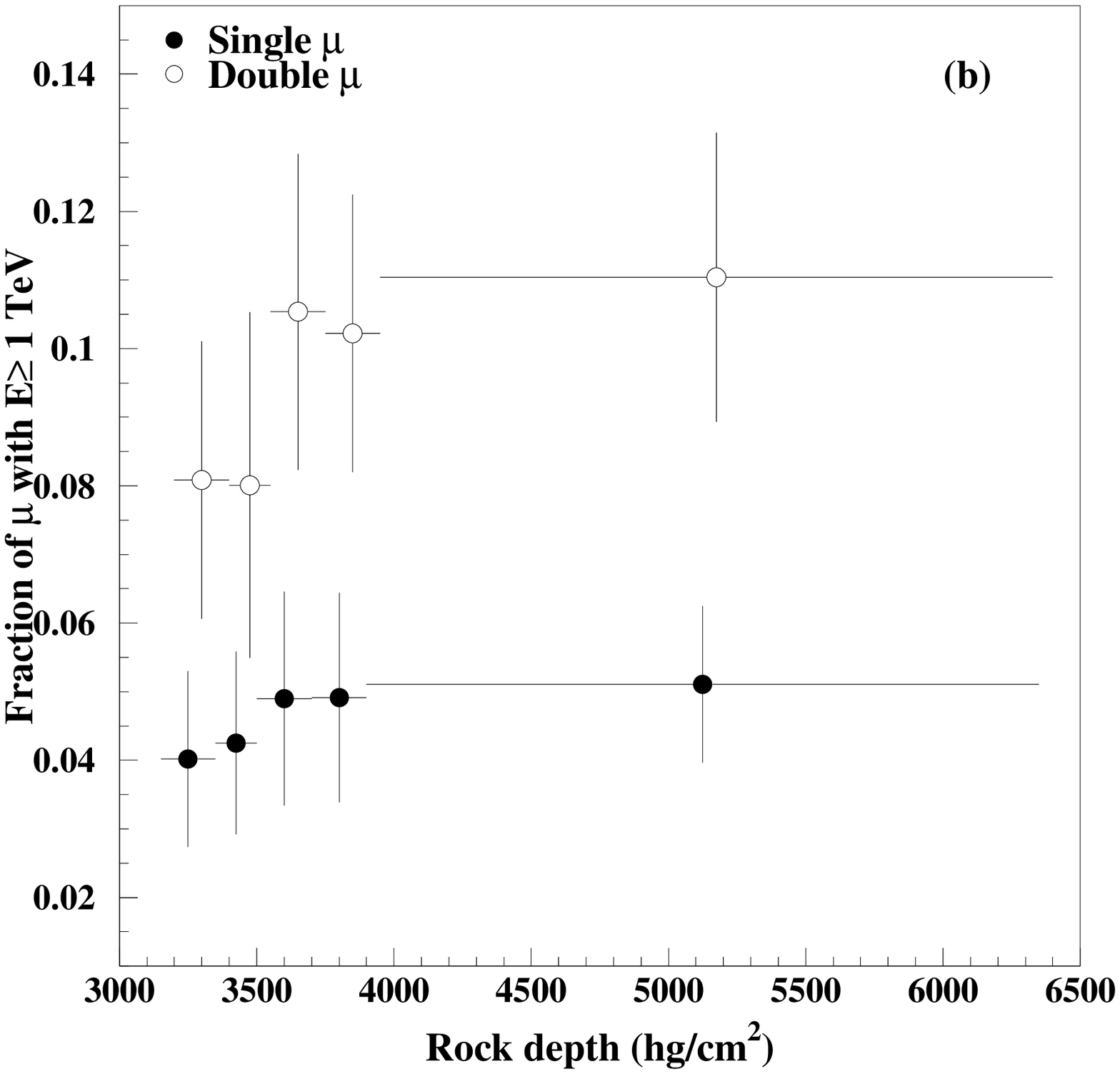,width=8.cm,height=6.5cm}
\end{tabular}
}},%
{(a) Average muon energy computed with a cut at $1~TeV$; (b)
fraction of muons with energies greater than $1~TeV$
versus the standard rock depth.}
\label{ecut}
\vspace{0.2cm}]
The TRD response is saturated for $E_{\mu} \geq
1~TeV$; for energies larger than $1~TeV$ only the number of events can 
be evaluated, while below $1~TeV$ we can reconstruct the energy distribution
and we can compute the average value cut to $1~TeV$.
\end{figwindow}

In Fig.~\ref{ecut}a the average energies of single muons (black circles) and of
double muons (open circles) for muons with $E_{\mu} < 1~TeV$ are
plotted versus the standard rock depth.
Fig.~\ref{ecut}b shows the fraction of events
with energy greater than $1~TeV$ versus the standard rock depth.
The values quoted in these figures have been obtained by combining 
the results coming from the three TRD modules. 
The error bars include statistical and estimated systematic uncertainties. 

The average muon energy for single muons with $E_{\mu} < 1~TeV$
is 
$196 \pm 3~ (stat) \pm 25~ (syst)~GeV$;
for double muons it is
$247 \pm 13~ (stat) \pm 25~ (syst)~GeV$.
The fraction of single muon
events with energies greater than $1~TeV$ is 
$4.6 \pm 0.1 ~(stat) \pm 1.4~(syst)~\%$,
while for the double muon events it is 
$9.4 \pm 0.6~ (stat) \pm 1.4~(syst)~\%$.

The experimental average muon energy over all energies was 
calculated by adding to the 
average energy obtained with an energy cut at $1~TeV$ the 
contribution from muons of greater energy. The high energy 
contribution was estimated by 
multiplying the measured fraction
of muons with energy $\geq 1~TeV$ by the average muon energy 
above $1~TeV$:
\begin{equation}
\label{eq:e3}
{<}E_{\mu}{>} = (1-f) \cdot {<}E_{\mu}{>}_{cut} + 
		f \cdot {<}E_{\mu}{>}_{no cut}
\end{equation}
where $f$ is the fraction of events with $E \geq 1~TeV$ 
(measured),
${<}E{>}_{cut}$ is the average energy with $E < 1~TeV$ 
(measured) and
${<}E{>}_{no cut}$ is the average energy with $E \geq 1~TeV$ (calculated).

\begin{figwindow}[1,r,%
{\mbox{
\epsfig{file=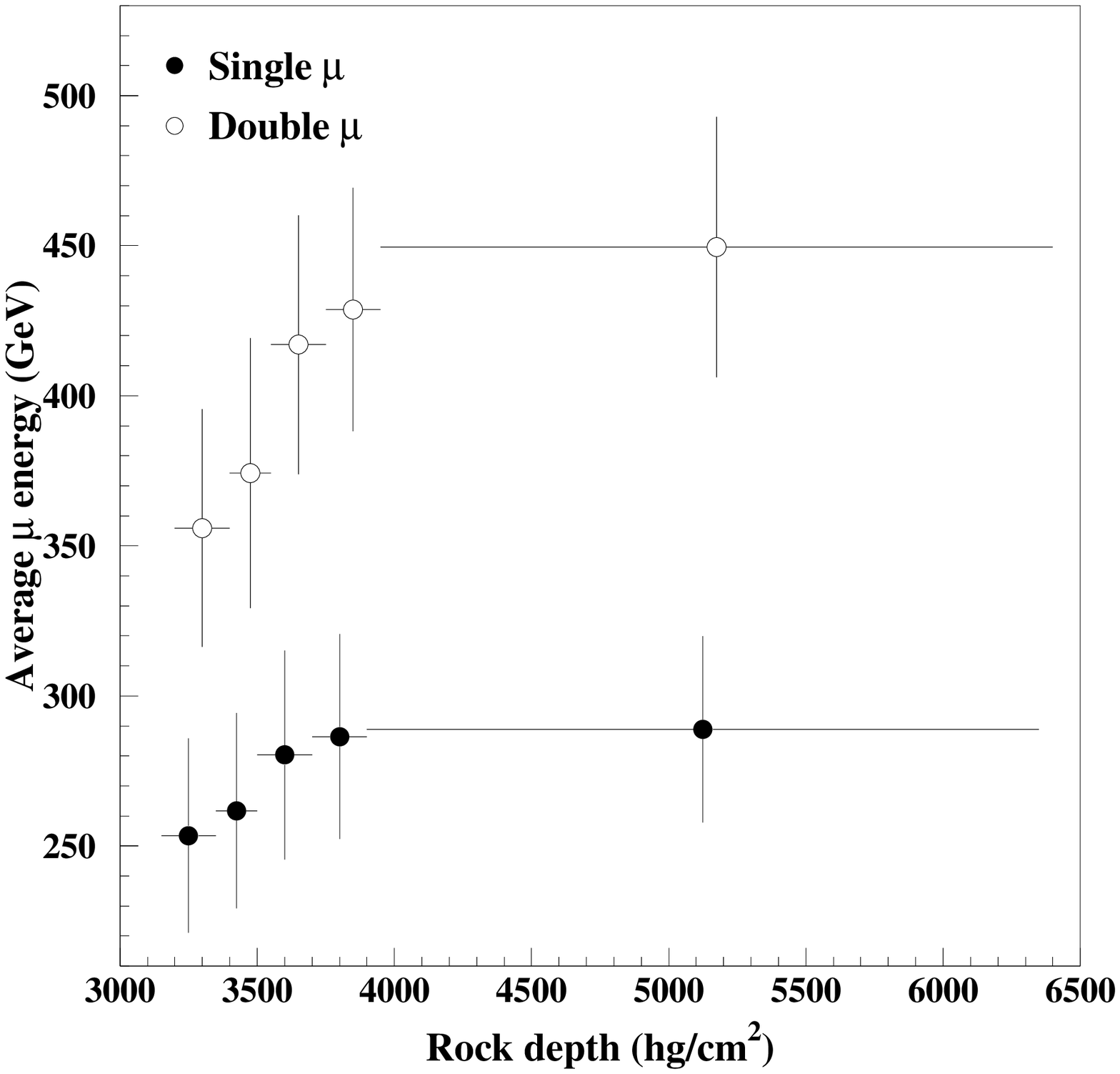,width=8.cm,height=7.0cm}}},%
{Average muon energy versus the standard rock depth.}
\label{etot}]
The evaluation of ${<}E{>}_{no cut}$ was based on a simple extrapolation 
of the local energy spectrum as reported in Equation (\ref{eq:e2}) using the
same parameters for the depth interval shown in Fig. \ref{ecut}. 
The average single muon 
energy obtained in this way is 
$272 \pm 4~(stat) \pm 33~(syst)~GeV$,
while for the double muons it is 
$398 \pm 16~(stat) \pm 39~(syst)~GeV$.

$~~~~~$These values of average muon energies do not change appreciably with
variations of $\beta$, $\epsilon$, and $\alpha$.  Varying each of these
three parameters by 3\%,
which is a value typically quoted (Battistoni
1997), results in a change in the average energies of 0.1\%, 0.2\%, and 1\%
respectively. These uncertainties are significantly smaller than our quoted
error.

$~~~~~$ Fig.~\ref{etot} shows the average for single muon energies 
(black circles)
and for double muons (open circles) as a function of
rock depth. The double muons are more energetic than single
ones, as predicted by the interaction models of primary cosmic rays with
the atmosphere.
\end{figwindow}
%
%
\vspace{1ex}
\begin{center}
{\Large\bf References}
\end{center}
%
Ambrosio, M., Antolini, R., Aramo, C., et al., 1999, Astroparticle
Physics 10, 11 \\
Barbarito, E., Bellotti, R., Cafagna, F. et al., 1995,
Nucl. Instr. Meth. A 365, 214 \\
D'Agostini, G., Nucl. Instr. Meth., 1995, A 362, 487\\
Mazziotta, M. N., 1995, LNGS 95/52 \\
Brun, R. et al., 1992, CERN Public. DD/EE/84-1\\
Lipari, P. and Stanev, T., 1991, Phys. Rev. D 44, 3543 \\
Battistoni, G., Ferrari, A., Forti, C. and Scapparone, E., 1997,
Nucl. Instr. Meth., A 394, 136
\end{document}